\documentclass{aa}
\usepackage{graphicx}
\usepackage{txfonts}
\usepackage{natbib}
\bibpunct{(}{)}{;}{a}{}{,}
\begin{document}
   \title{First polarimetric observations and modeling of the FeH F$^{4}\Delta$--X$^{4}\Delta$ system}

   \author{N. Afram\inst{1}
          \and S.~V.~Berdyugina\inst{1,2}
          \and D.~M.~Fluri\inst{1}
          \and M.~Semel\inst{3}
          \and M.~Bianda\inst{1,4}
          \and R.~Ramelli\inst{4}
          }

   \institute{Institute of Astronomy, ETH Zurich, 8092 Zurich, Switzerland\\
   \email{nafram@astro.phys.ethz.ch}
   \and       
           Tuorla Observatory, University of Turku, 21500 Pikki\"o, Finland
   \and
           Laboratoire d'Etudes Spatiales et d'Instrumentation en Astrophysique, Observatoire de Paris, Section de Meudon, F-92195 Meudon Principal Cedex, France
   \and       Istituto Ricerche Solari Locarno, Via Patocchi, 6605 Locarno-Monti, Switzerland
           }    
   \date{Received date; accepted date}

\abstract
{Lines of diatomic molecules are typically much more temperature and pressure sensitive than atomic lines, which makes them ideal, complementary tools for studying cool stellar atmospheres as well as the internal structure of sunspots and starspots. The FeH F$^4\Delta\!-–\!$X$^4\Delta$ system represents such an example that exhibits in addition a large magnetic field sensitivity. However, the current theoretical descriptions of these transitions including the molecular constants involved are only based on intensity measurements because polarimetric observations have not been available so far, which limits their diagnostic value. Furthermore, the theory was optimized to reproduce energy levels and line strengths without taking the magnetic sensitivities into account.}
{We present for the first time spectropolarimetric observations of the FeH F$^4\Delta\!-–\!$X$^4\Delta$ system measured in sunspots to investigate their diagnostic capabilities for probing solar and stellar magnetic fields. In particular, we investigate whether the current theoretical model of FeH can reproduce the observed Stokes profiles including their magnetic properties.}
{The polarimetric observations of the FeH F$^4\Delta\!-–\!$X$^4\Delta$ system in Stokes $I$ and $V$ are compared with synthetic Stokes profiles modeled with radiative transfer calculations. This allows us to infer the temperature and the magnetic field strength of the observed sunspots.}
{We find that the current theory  successfully reproduces the magnetic properties of a large number of lines in the FeH F$^4\Delta\!-–\!$X$^4\Delta$ system. In a few cases the observations indicate a larger Zeeman splitting than predicted by the theory. There, our observations have provided additional constraints, which allowed us to determine empirical molecular constants.}
{The FeH F$^4\Delta\!-–\!$X$^4\Delta$ system is found to be a very sensitive magnetic diagnostic tool. Polarimetric data of these lines, in contrast to intensity measurements, provide us with more direct and detailed information to study the coolest parts of sunspot and starspot umbrae, and cool active dwarfs.}


   \maketitle

\section{Introduction}

  Many lines of diatomic molecules are not only good temperature and pressure indicators but also excellent magnetic field diagnostics, as has been shown by \citet{berdyuginaetal2000} and \citet{berdyuginasolanki2002} in their overview of magnetic properties of molecules observed in spectra of sunspots and cool stars. 

Iron hydride (FeH) is one of the most sensitive indicators of magnetic fields in cool atmospheres, but so far observations have been restricted to intensity spectra. \citet{wallaceetal1998} revealed the remarkable magnetic sensitivity of many lines of the FeH  F$^{4}\Delta$--X$^{4}\Delta$ system in sunspot observations. \citet{valentietal2001} detected Zeeman broadening of FeH lines in an active M dwarf and illustrated the usefulness of these lines in studying stellar magnetic fields. Because of the lack of a theoretical description of the FeH molecule, they modeled the stellar spectrum employing the sunspot measurements by \citet{wallaceetal1998}. First modeling of synthetic Stokes profiles of FeH lines indicated  that this particular electronic system will be a powerful tool for diagnosing solar and stellar magnetism once the spin-coupling constants become available \citep{berdyuginaetal2003}. 

The set of missing molecular constants was provided by \citet{dulicketal2003}. They determined the spin-orbit ($A$) and spin-spin ($\lambda$) constants for the FeH  F$^{4}\Delta$--X$^{4}\Delta$ system, using the laboratory measurements by \citet{phillipsetal1987}. They found that the spin splittings for the observed vibrational levels could be reproduced within an error of 0.5 cm$^{-1}$ by a single pair of $A$ and $\lambda$ values for each electronic state. However, the analysis by \citet{dulicketal2003} aimed at modeling energy levels and line strengths without considering magnetic characteristics. Only if the theory correctly describes the magnetic properties we can exploit the full diagnostic value of FeH.

Here we present first polarimetric measurements of the FeH  F$^{4}\Delta$--X$^{4}\Delta$ system which provide novel constraints and tests for the theory. They were recorded in a sunspot and are described in Sect.~\ref{obs}.  We carried out a perturbation analysis for the FeH F$^{4}\Delta$--X$^{4}\Delta$ system  to account for the Zeeman effect as described 
  in Sect.~\ref{theory}. Molecular Stokes line profiles in the presence of magnetic fields  are calculated and compared with the sunspot observations in Sect.~\ref{sto}. Our results are based on the molecular constants determined by \citet{dulicketal2003} to test the current theory.  We conclude that  polarization measurements of the considered FeH system and their successful theoretical interpretation strongly constrain molecular parameters and greatly advance our ability to explore magnetic fields in cool stellar atmospheres.

\section{Observations}\label{obs}
\noindent
Observations were performed at Istituto Ricerche Solari Locarno
(IRSOL) with the 45 cm aperture Gregory Coud\'e telescope, the high
spectral resolution Czerny Turner spectrograph, and a two beam
exchange polarimeter.
On 23rd August 2004 registrations were done in a sunspot (NOAA 0663).
Wavelength intervals of about 5 \AA\ were observed around 9902 \AA, 9906 \AA, 9976 \AA, 9979 \AA,  10059 \AA, and 10062 \AA.

The two beam exchange technique by \citet{semeletal1993} had already
been successfully applied at IRSOL for polarimetric observations in the
blue part of the spectrum \citep{biandaetal1998}.
To observe the FeH lines in the near infrared we used the original
Semel polarimeter described by \citet{semeletal2001} without
collimation lenses.

The polarization analyzer consists of a set of motorized Fresnel
Rhomb retarders installed in front of a Savart Plate beam splitter.
The retarders are rotated in order to separate the incoming beam into
two beams corresponding to $I + X$ and $I - X$ with $X$ being Stokes
$Q$,  $U$, or  $V$.
Thus, two images of opposite polarization of the same solar region were
projected on the spectrograph slit jaws. To obtain a single Stokes measurement two such spectral recordings were
registered with a CCD subsequently changing between the polarization state of
the two beams with a suitable rotation of the retarders. On the spectrograph slit, two images with orthogonal polarization are projected, each subtending $35 \arcsec$ in the spatial direction along the slit direction. The position on the spectrograph slit was chosen in order to maximize the length of the segment covering the sunspot umbra. The profiles in Fig. 1-3 are obtained averaging the coolest part of the umbra spectrum over $7 \arcsec$.
The exposure time for each frame was 5 seconds. A complete cycle
of Stokes vector observation required six frames and took about 2
minutes, leading to a spatial resolution of 2 arcsec. Two cycles were performed for
each observed spectral interval.

Dark frames were recorded at the beginning and at the end of the
observation run.
Flat field observations were performed near the solar disk center on
quiet regions. Details of the data reduction technique are described by
\citet{biandaetal1998}. To correct for the instrumental polarization
the polarization values obtained in the flat field registrations at
disk center were subtracted from those measured in the sunspots.

\section{Theoretical approach}\label{theory}

\noindent If a molecule possesses a nonzero magnetic moment, it
interacts with an external magnetic field and causes a precession of
the total angular momentum $\vec{J}$ about the field direction. As a
result one can observe the Zeeman effect, i.e. the magnetic splitting of energy
levels in the molecule.

The energies of magnetic components depend on how the
angular momenta are coupled to the rotation of the molecule.
\noindent The FeH F$^{4}\Delta$--X$^{4}\Delta$ system is produced by
transitions between two electronic states with couplings of the
angular momenta that are intermediate between the limiting Hund's cases
(a) and (b). The perturbation calculation of the molecular Zeeman
effect is based on the effective Hamiltonian for this transition, which can be represented by two parts, $\mathcal{H}^{ab}$
and $\mathcal{H}^{H}$,

\begin{equation}\label{eq:H}
\mathcal{H} =\mathcal{H}^{ab}+\mathcal{H}^{H},
\end{equation}
 the former performing the transformation of the Hund's
case (a) wavefunctions  into the intermediate case wavefunctions  and the latter describing the
interaction with the external magnetic field. The matrix elements of the Hamiltonian $\mathcal{H}^{ab}$ include the spin-orbital,
spin-spin, rotational, and centrifugal distortion interactions in the electronic state described
with Hund's case (a) wavefunctions as a basis set \citep{dulicketal2003}.
 By diagonalization of the total Hamiltonian  $\mathcal{H}$ we obtain the energy levels, the transition strengths, and the Zeeman pattern of the magnetically perturbed system \citep[see][]{berdyuginaetal2005}.

\section{Stokes profiles in the Zeeman regime}\label{sto}

\noindent By implementing the Hamiltonian given in Eq.~(\ref{eq:H}), we can now calculate molecular line profiles in the presence of magnetic fields in the Zeeman regime, using the radiative transfer code STOPRO \citep{solanki1987,solankietal1992,frutigeretal2000,berdyuginaetal2003,berdyuginaetal2005}. These synthetic Stokes profiles are compared with profiles observed in sunspots. For this purpose we vary only the model atmosphere and the magnetic field  in our fitting procedure.  Atmospheric  models are taken from \citet{hauschildtetal1999}.  To account for the photospheric stray light, we combined calculated sunspot and photospheric spectra using various spot filling factors.



\begin{figure}
\centering
\resizebox{\hsize}{!}{\includegraphics{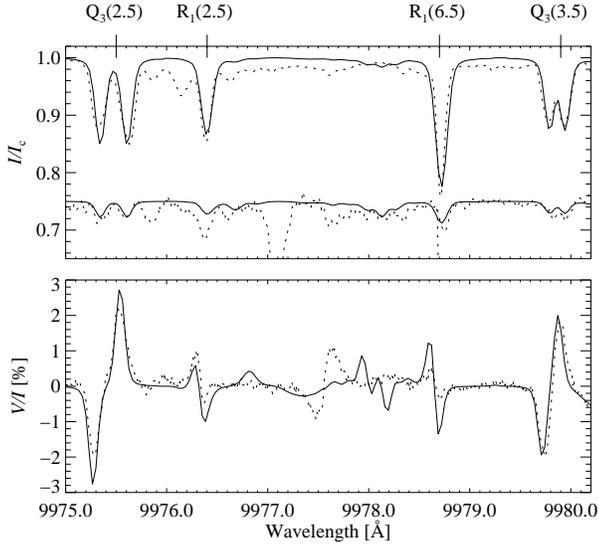}}
\caption{FeH Stokes $I$ and Stokes $V$ line profiles (upper and lower
  panel, respectively). Solid lines represent synthetic Stokes
  profiles; observed Stokes $I$ and $V$
  profiles are shown with dotted lines. The top Stokes $I$ spectrum shows a sunspot
  observation from \citet{wallaceetal1998}, while the bottom
  Stokes $I$ spectrum (shifted for the sake of clarity) as well as the Stokes $V$ spectrum correspond to  our own sunspot observations at IRSOL. Synthetic profiles are based on the theory by \citet{dulicketal2003} and were calculated with a field strength of 3~kG, umbral models with effective temperatures $T_{\rm{eff}}$ of 4300~K and 4800~K \citep{hauschildtetal1999}, and  filling factors of
  0.95 and 0.8 to fit the Wallace and IRSOL data, respectively.}
\label{fig:semdul1}
\end{figure}

With the  molecular constants from  \citet{dulicketal2003} many spectral lines with various $\Delta J$, $J$, and $\Omega$ can be reproduced, $J$ being the total angular momentum and $\Omega$ the quantum number for the component of the total electronic angular momentum along the internuclear axis of the molecule. 

In Figs.~1 and 2 we present such examples of observed FeH Stokes $I$ and $V$ profiles compared with the calculated spectra. Stokes $I$ observations are taken from \citet{wallaceetal1998} and IRSOL (upper panels), whereas Stokes $V$ spectra (lower panels) were obtained only and for the first time in observations at IRSOL. The fact that the spots observed at IRSOL were neither very large nor cold is reflected in the shallow depth of the FeH intensity profiles as compared to the spectra from the sunspot atlas by \citet{wallaceetal1998}. This leads to smaller filling factors in the IRSOL spectra since the influence of stray light from the photosphere becomes more relevant. Nonetheless, the Stokes $V$ profiles of the FeH lines have very clear, easily measurable signals with a degree of circular polarization of the order of 1\% to 3\%. The magnetic field strength was found to have the same value of 3~kG in both the \citet{wallaceetal1998} and the IRSOL data sets. In both cases the magnetic field vector was assumed to be aligned along the line of sight in our modeling, pointing away from the observer, although this direction can of course only be determined for the IRSOL data thanks to the spectropolarimetric measurements. Since FeH lines effectively form at log $\tau(1\mu)=-2.5$, according to our model, the inferred magnetic field corresponds to this depth.

\begin{figure}
\centering
\resizebox{\hsize}{!}{\includegraphics{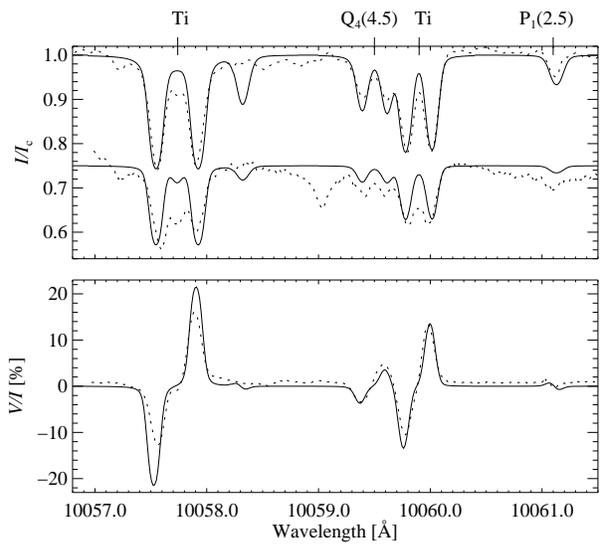}}
\caption{The same as Fig.~\ref{fig:semdul1} but for a different wavelength region. Synthetic profiles (solid) are based on the theory by \citet{dulicketal2003} and were calculated with a field strength of 3~kG, umbral models with $T_{\rm{eff}}$ of 4000~K and 4500~K \citep{hauschildtetal1999}, and  filling factors of 0.97 and 0.8 to fit the Wallace and IRSOL data, respectively.}
\label{fig:semdul2}
\end{figure}

The main features of both Stokes $I$ and $V$ could be reproduced, suggesting that our calculations of the molecular Zeeman splitting in these FeH  lines based on the theory by \citet{dulicketal2003} are realistic. Lines of the $Q$-branch  with  $J\le4.5$ exhibit particularly strong magnetic sensitivity  as was expected from earlier investigations  \citep{berdyuginasolanki2002}. Also the $P$ and $R$ lines  with  $\Omega\lesssim2.5$ and $J\lesssim10.5$, such as those presented in Figs.~\ref{fig:semdul1} and \ref{fig:semdul2}, can be properly modeled with the molecular constants from \citet{dulicketal2003}. Note in particular the opposite sign of the Stokes $V$ profiles in the $P$- and $R$-lines as compared with the $Q$-lines, which is due to the opposite sign of the effective Land\'e factors. This represents an example of important additional constraints for the theory that remain inaccessible by intensity measurements and can only be provided by spectropolarimetric observations.

\begin{figure}
\centering
\resizebox{\hsize}{!}{\includegraphics{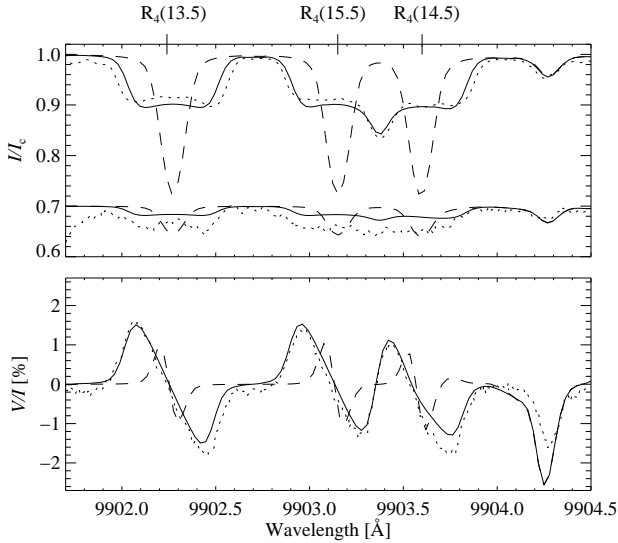}}
\caption{The same as Fig.~\ref{fig:semdul1} for a different spectral window containing FeH lines that suffer from unknown perturbations. Observations (dotted) are compared with two different models, the first being based on our empirically determined spin-orbit constants (solid), and the second one (dashed) assuming the theory by \citet{dulicketal2003}. Both synthetic spectra were calculated with a field strength of 3~kG, umbral models with $T_{\rm{eff}}$ of 4300~K and 4800~K \citep{hauschildtetal1999}, and filling factors of
   0.98 and 0.9 to fit the Wallace and IRSOL data, respectively.}
\label{fig:semnodul}
\end{figure}

Some of the observed line profiles arising from the FeH  F$^{4}\Delta$--X$^{4}\Delta$ system cannot be modeled correctly when employing the theory by \citet{dulicketal2003}, e.g.\ the $R_4$  lines\footnote{Here we denote rotational branches with indices 1, 2, 3, and 4 as corresponding to transitions between levels both having $\Omega=$ $1/2$, $3/2$, $5/2$, and $7/2$.} in the region from 9900~\AA\ to 10000~\AA.  Figure \ref{fig:semnodul} illustrates examples of such $R_4$ lines which would be ideally suited for magnetic field diagnostics thanks to their strong splitting and peculiar profile shapes.  However, their effective Land\'e factors and thus their Zeeman splitting as calculated with the current theory are about one order of magnitude too small. These lines with a large $\Omega=\!3.5$   and  $J\gtrsim10.5$ seem to be affected by unknown perturbations neglected in the theory such that neither the intensity nor the polarization profiles are  properly reproduced. For those transitions we searched for ideal spin-orbit  interaction constants $A$ with a $\chi^2$ minimization.  Best fits to the Stokes profiles were obtained for $A''=-30$\,cm$^{-1}$ and $A'=-330$\,cm$^{-1}$, for the lower and upper level respectively.

 Unknown perturbations appear to have a small influence on lines of the $Q$-branch ($\Delta J\!=\!0$) in which the $J$-number of the lower and upper level coincide so that perturbations seemingly cancel out. For the $P$- and $R$-branches such cancelations do not occur and perturbations might become a major issue for lines  with  $\Omega\gtrsim2.5$ and $J\gtrsim10.5$, which will be investigated in a forthcoming paper. Such a behavior is similar to that discussed for the TiO $\gamma'$ system by \citet{berdyuginasolanki2002}.

\section{Conclusions}
   \noindent The first polarimetric observations of the FeH F$^{4}\Delta$--X$^{4}\Delta$ system and their successful modeling open an opportunity to fully exploit the diagnostic
capabilities of this molecular system, which represents a
powerful tool for solar and stellar magnetic field studies.

We have described the synthesis of molecular Stokes
parameters and showed that  it was possible to fit a wide range of available  spectral lines in different wavelength regions with various $\Delta J$, $J$, and $\Omega$ numbers with the set of molecular constants provided by  \citet{dulicketal2003}, especially lines with small $J$-numbers.  Our analysis has demonstrated that the theory produces not only the true Zeeman splitting but also the right sign for the Stokes $V$ profiles. Since the theory for these lines seems to be understood, they represent good candidates for determining the thermal and magnetic structure of cool atmospheres.

Nevertheless, interesting spectral regions exist, such as the region around 9900\,\AA\,, where a strong splitting is observed that cannot be modeled with the available constants.  It seems that an unknown  perturbation, which is not taken into account by the applied Hamiltonian, affects some of the magnetically very sensitive lines.  However, using spin-orbit constants which were determined empirically, we were able to reproduce remarkably strongly split Stokes
$I$ and Stokes $V$ profiles observed in a sunspot spectrum, such as the  $R_4(13.5)$, $R_4(14.5)$, and $R_4(15.5)$ lines. 

We conclude that our ability to model the FeH F$^{4}\Delta$--X$^{4}\Delta$ system based on a set of theoretical and empirical molecular constants provides a new valuable source of information about the physical parameters in observed objects.


\begin{acknowledgements}
We acknowledge
the EURYI award from the ESF, the SNF grants PE002-104552
and 200021-103696, and the ETH Research Grant TH-2/04-3.
The IRSOL obtains financial support provided by the canton of Ticino, the city of Locarno, ETH Zurich, and the Fondazione Aldo and Cele Dacc\`o.
\end{acknowledgements}


\bibliographystyle{aa}
\bibliography{journals,nafram}

\begin{thebibliography}{15}
\expandafter\ifx\csname natexlab\endcsname\relax\def\natexlab#1{#1}\fi

\bibitem[{{Berdyugina} {et~al.}(2005){Berdyugina}, {Braun}, {Fluri}, \&
  {Solanki}}]{berdyuginaetal2005}
{Berdyugina}, S.~V., {Braun}, P.~A., {Fluri}, D.~M., \& {Solanki}, S.~K. 2005,
  \aap, 444, 947

\bibitem[{{Berdyugina} {et~al.}(2000){Berdyugina}, {Frutiger}, {Solanki}, \&
  {Livingston}}]{berdyuginaetal2000}
{Berdyugina}, S.~V., {Frutiger}, C., {Solanki}, S.~K., \& {Livingston}, W.
  2000, {\aap}, 364, L101

\bibitem[{{Berdyugina} \& {Solanki}(2002)}]{berdyuginasolanki2002}
{Berdyugina}, S.~V. \& {Solanki}, S.~K. 2002, \aap, 385, 701

\bibitem[{{Berdyugina} {et~al.}(2003){Berdyugina}, {Solanki}, \&
  {Frutiger}}]{berdyuginaetal2003}
{Berdyugina}, S.~V., {Solanki}, S.~K., \& {Frutiger}, C. 2003, \aap, 412, 513

\bibitem[{{Bianda} {et~al.}(1998){Bianda}, {Solanki}, \&
  {Stenflo}}]{biandaetal1998}
{Bianda}, M., {Solanki}, S.~K., \& {Stenflo}, J.~O. 1998, \aap, 331, 760

\bibitem[{{Dulick} {et~al.}(2003){Dulick}, {Bauschlicher Jr.}, {Burrows},
  {Sharp}, {Ram}, \& {Bernath}}]{dulicketal2003}
{Dulick}, M., {Bauschlicher Jr.}, C., {Burrows}, A., {et~al.} 2003, ApJ, 594,
  651

\bibitem[{{Frutiger} {et~al.}(2000){Frutiger}, {Solanki}, {Fligge}, \&
  {Bruls}}]{frutigeretal2000}
{Frutiger}, C., {Solanki}, S.~K., {Fligge}, M., \& {Bruls}, J. H. M.~J. 2000,
  \aap, 358, 1109

\bibitem[{{Hauschildt} {et~al.}(1999){Hauschildt}, {Allard}, \&
  {Baron}}]{hauschildtetal1999}
{Hauschildt}, P.~H., {Allard}, F., \& {Baron}, E. 1999, ApJ, 512, 377

\bibitem[{{Phillips} {et~al.}(1987){Phillips}, {Davis}, {Lindgren}, \&
  {Balfour}}]{phillipsetal1987}
{Phillips}, J.~G., {Davis}, S., {Lindgren}, B., \& {Balfour}, W.~J. 1987, ApJS,
  65, 721

\bibitem[{{Semel} {et~al.}(1993){Semel}, {Donati}, \& {Rees}}]{semeletal1993}
{Semel}, M., {Donati}, J.-F., \& {Rees}, D.~E. 1993, \aap, 278, 231

\bibitem[{{Semel} \& {L{\'o}pez Ariste}(2001)}]{semeletal2001}
{Semel}, M. \& {L{\'o}pez Ariste}, A. 2001, in ASP Conf. Ser., Vol. 248,
  Magnetic Fields Across the Hertzsprung-Russell Diagram, ed. G.~{Mathys},
  S.~K. {Solanki}, \& D.~T. {Wickramasinghe}, 575

\bibitem[{{Solanki}(1987)}]{solanki1987}
{Solanki}, S.~K. 1987, PhD thesis, ETH, Zurich, Switzerland

\bibitem[{{Solanki} {et~al.}(1992){Solanki}, {R\"uedi}, \&
  {Livingston}}]{solankietal1992}
{Solanki}, S.~K., {R\"uedi}, I., \& {Livingston}, W. 1992, \aap, 263, 312

\bibitem[{{Valenti} {et~al.}(2001){Valenti}, {Johns-Krull}, \&
  {Piskunov}}]{valentietal2001}
{Valenti}, J.~A., {Johns-Krull}, C.~M., \& {Piskunov}, N.~E. 2001, in ASP Conf.
  Ser., Vol. 223, 11th Cool Stars, Stellar Systems and the Sun, ed. R.~J.
  {Garc\'{\i}a L\'opez}, R.~{Rebolo}, \& M.~R. {Zapatero Osorio}, 1579

\bibitem[{{Wallace} {et~al.}(1998){Wallace}, {Livingston}, {Bernath}, \&
  {Ram}}]{wallaceetal1998}
{Wallace}, L., {Livingston}, W.~C., {Bernath}, P.~F., \& {Ram}, R.~S. 1998, An
  Atlas of the Sunspot Umbral Spectrum in the Red and Infrared from $8900$ to
  $15 050$ cm$^{-1}$ ($6642$ to $11 230$ \AA) (NOAO)

\end{thebibliography}
\end{document}